\documentclass[]{article}
\usepackage{graphicx}
\usepackage{fancyhdr}
\usepackage{multicol}
\usepackage{styl-ap}

\begin{document}
\title{Analysis of the  MACRO experiment data to compare particles arrival times under Gran Sasso}
\author{Francesco Ronga\work{1}}
\workplace{INFN Laboratori Nazionali di Frascati, Frascati Italy}

\mainauthor{francesco.ronga@lnf.infn.it}
\maketitle

\begin{abstract}%
The claim of a neutrino velocity different from the speed of the light, made in September 2001 by the Opera experiment,  suggested the study of the time delays between TeV underground muons in the Gran Sasso laboratory using the old data of the MACRO experiment, ended in 2000. This study can give also hints on new physics in the particle cascade produced by the interaction of a cosmic ray with the atmosphere.
\end{abstract}

\keywords{neutrino velocity - new massive particles - tachyons - supersymmetry}

\section{Introduction}

In September 2011 there was a measurement  of the  speed of neutrino faster than the speed of light  by $\frac{v-c}{c} = 2.48 \pm 0.28(stat)\pm 0.30(sys) \times 10^{-5}$ (Adam, 2011). After many checks we know now that this result was due to hardware problems and  the Opera 2012 result is that the speed of the neutrinos traveling from CERN to the Gran Sasso is $\frac{v-c}{c} = -0.7 \pm 0.5(stat) ^{+2.5}_{-1.5}(sys) \times 10^{-6}$ (Dracos,  2012). This result is in agreement with the results of the other Gran Sasso experiments (Bertolucci,  2012).

However the interest for this claim suggested the possibility to compare neutrino and muon velocity in a cosmic ray cascade.
(Montaruli and Ronga, 2011). The interaction of a primary cosmic ray  with the atmosphere produces a cascade with many kind of particles, and in particular neutrinos and muons. Muon neutrinos and muons are produced  mainly via the decay of charged pions and kaons produced in the primary cosmic ray interactions. Above about 10 TeV they can come also from prompt decays of charmed hadrons. This  component has not yet been observed. In a deep underground detector only muon and neutrino are detected. 
If the neutrino velocity is different from $c$ the  neutrinos in this cascade,  should arrive with times  different  from the times of the muons from the same parent decay, or from another decay, with a time delay that should change according to the neutrino path length that depends on its zenith angle $\theta$. In underground detectors muon neutrinos are detected looking for induced muons produced by neutrino charged current interactions in the rock, or in  the ice around or inside the instrumented region. Hence, a time spread should be observed between the muons produced  directly by the pion or kaon decay  and the muon produced by neutrino interactions.

The  path length from the meson decay point is a few tens of kilometers for vertical neutrinos  and up to $\sim 300$~km for near horizontal neutrinos. 
Assuming the original  time difference  observed in OPERA  nearly horizontal neutrinos should arrive up to 28 nsec before the other secondaries. In  (Gaisser  and Stanev,1998) a table of average production heights neutrinos in the atmosphere has been reported. The typical production height for neutrinos of energy above 20 GeV can be 17.6 km at the vertical, 94.9 km at $\cos\theta = 0.25$ and 335.7 km at $\cos\theta = 0.05$, which would correspond to 1.4, 78 and 27.6~ns. 


There are already limits of tachions or anomalous delayed particles in cosmic rays. The limits are  obtained searching for example signals before  or after the main front of the electromagnetic shower. But this kind of searches stopped some time ago and the last particle data book  review of those data is the one of 1994 (Montanet  et al, 1994). The  limits obtained are of small interest in the framework of the OPERA result. However, if neutrinos were tachions, it is likely that other kind of tachions could exist and this search in very high energy cosmic rays could have a new interest. 

It is important to note that the Gran Sasso mountain minimum depth $\sim 2700~gr/cm^2$ correspond to a minimum muon energy of  1.4 TeV.
It easy to compute that, requiring a minimum threshold of 50 MeV in the detector the time difference between two muons underground should be $\ll ~0.2~nsec$. Therefore anomalous time differences should be a signal of "new physics", for example  signal of a supersymmetric massive particles produced in a cosmic ray cascade. For example, let we assume an hypothetical hadron of mass 100 GeV, produced by a an interaction of a proton with center of mass energy 7 TeV (the LHC energy). If this hypothetical hadron interacts or decays  after 10 Km producing at the end muons, the delay between mountain muon from the massive particle and the muon produced in the primary vertex is of the order of 13 nsec.
LHC experiments have put limits for new  hadron-like massive particles (Chatrchyan, 2012 and  Aad,  2012)  but it is important to remember that the cosmic ray energy could be larger than the LHC energy. Under Gran Sasso the fraction of multiple muons produced by cosmic ray with center of mass energy $\ge7~TeV$ is estimated of the order of $10^{-3}$ in MACRO, corresponding to several thousand multiple muon  events in the MACRO data set. 
One should also consider the possibility that new 
massive relic particles are directly in the primary cosmic radiations.

The MACRO experiment has done several searches for possible anomalies of the time differences between  muons (Ahlen et al, 1991).
 The search  was done mainly to study time differences of the order of a few msec or more, but this paper contains also the study of time differences at the ns-level.  
 The statistics was limited to 35832 tracks in events with two or more tracks. In 1992 none was thinking to  tachionic neutrinos and therefore there was no estimate of the number of tracks due to down-going neutrino together with  a primary muons. In (Scapparone,1995) this study was extended to  about 140000 tracks of multi muon events, corresponding to about $4\%$ of the total MACRO statistics. The time distribution was in agreement with the predictions.

 \begin{myfigure}
\centerline{\resizebox{130mm}{!}{\includegraphics{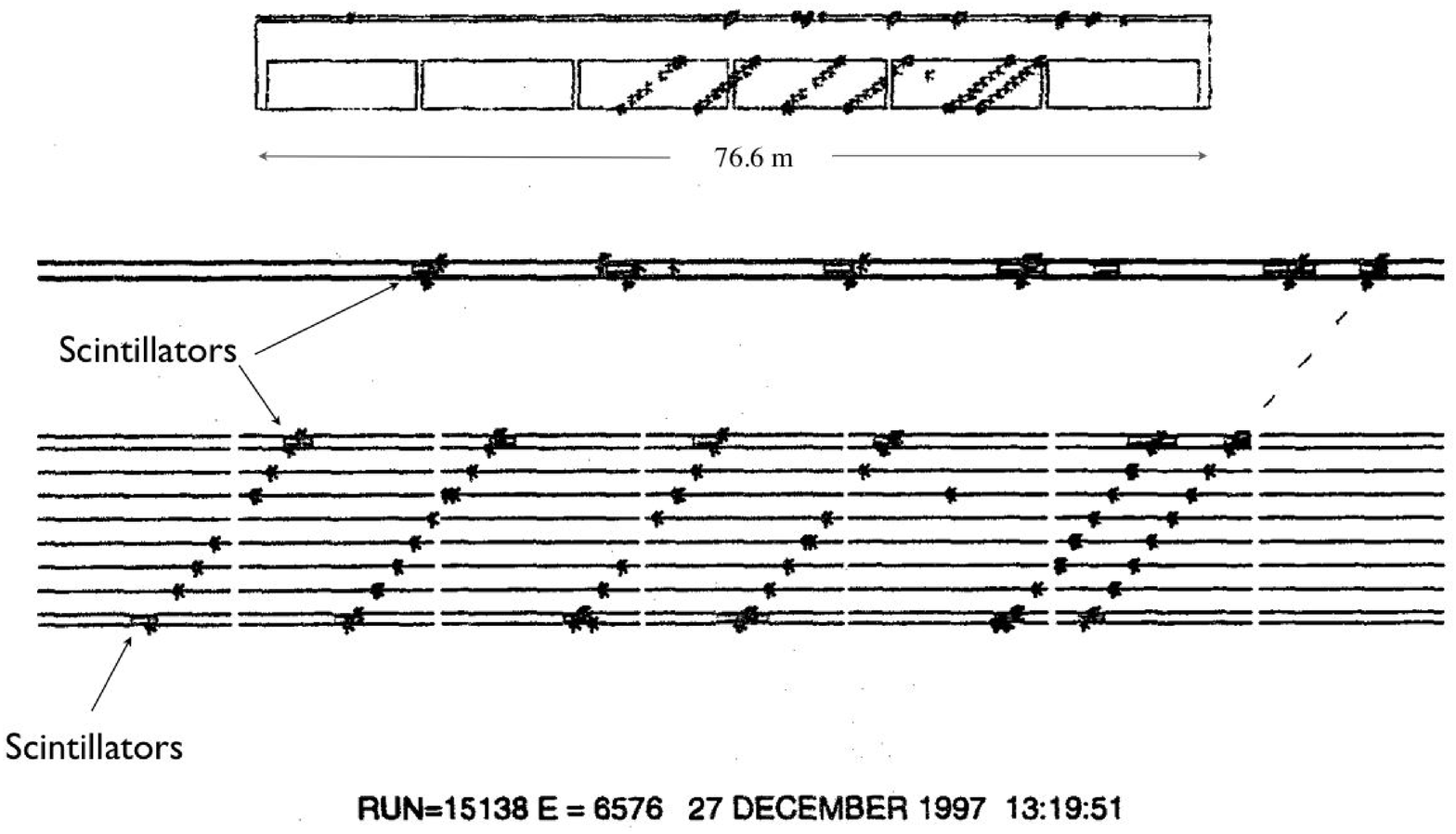}}}
\caption{Event with six parallel muons  in 3 MACRO "supermodules". On the top there is the full MACRO display, on the bottom the zoom of the 3 supermodules interested by the event. The 12 steamer tube horizontal planes are shown as horizontal lines, the back points are the streamer tubes fired; the scintillator boxes fired are shown as rectangles.}
\label{author-fig1}
\end{myfigure}

In this paper I present an analysis of the full MACRO statistics. This was not an easy job. The main reason is that MACRO ended in 2000 and that most of the analysis software was designed for VAX/AlphaVAX computers and data formats around 1990 (a geological era for computers!). A lot of time has been necessary to convert programs and to find data files, some time stored on data tape cassette of old formats, obsolete and not supported by modern computers.

\section{The MACRO experiment and the timing system}
The MACRO experiment (Ambrosio, 2002) was located in the Hall B of the Gran Sasso
underground laboratory. The modularity allowed the data-taking also with
partial configurations of the apparatus, starting since March 1989.
The full detector was operative in the period April 1994 - December
2000.

MACRO was a large rectangular box ($76.6 \times 12 \times 9.3 \
m^3$) divided longitudinally in 6 supermodules and vertically in
a lower and an upper part (called {\it attico}). The active
elements were liquid scintillator counters for time measurement
and streamer tubes for tracking, with $27^\circ$ stereo strip
readouts. The lower half of the detector was filled with trays of
crushed rock absorber alternating with streamer tube planes, while
the {\it attico} was hollow and contained the electronics racks
and work areas. The rock absorber sets a minimum energy threshold
for vertical muons of $1\ GeV$.

The tracking system has been designed to reconstruct the particle
trajectory
in different views ($x-z$ for horizontal streamer tubes, $d-z$
for horizontal strips, $y-z$ for vertical streamer tubes combined
with central hits).  To perform this analysis the standard MACRO tracking software has been improved
to have a larger efficiency for near horizontal tracks.


The intrinsic angular resolution for muons typically ranges from $0.2^\circ$
to $1^\circ$ depending on track length. This resolution is lower than the
angular spread due to multiple scattering of downward-going muons in the
rock.


The scintillator system consisted of horizontal and vertical layers of
counters filled with a mixture of mineral oil (96.4\%), pseudocumene
(3.6\%) and wavelength shifters ($2.88\ g/l$).
 The counters had an active volume of $11.2 \times 0.73 \times 0.19\ m^3$
 in the horizontal planes and $11.1 \times 0.22 \times 0.46\ m^3$ in the
 vertical planes.

The total charge and the time of occurrence of the signals were
measured at the two ends 
of each counter with 
two independent systems, the Energy Response Processor (ERP) and
the Pulse Height Recorder and Synchronous Encoder (PHRASE). 
The analysis
described in this paper are based on ERP data. 
Time and longitudinal position resolution for single muon in a counter
were about $0.6\ ns$ and $12\ cm$, 
respectively. The photomultiplier signal is split into a direct output
and one attenuated by a factor 10, in order to be on-scale also for 
very large pulses. Two different thresholds are used for the timing of
these two outputs. The redundancy of the time measurement helps to
eliminate spurious effects. Each MACRO supermodule is connected to a dedicated independent ERP system.
The timing between the ERP systems is insured by standard CAMAC TDC. Due to the random noise
the possibility to have wrong times in the inter ERP TDCs is quite high and this is the main  source of not gaussian
tails in the time distributions for events interesting different supermodules.

\begin{myfigure}
\centerline{\resizebox{90mm}{!}{\includegraphics{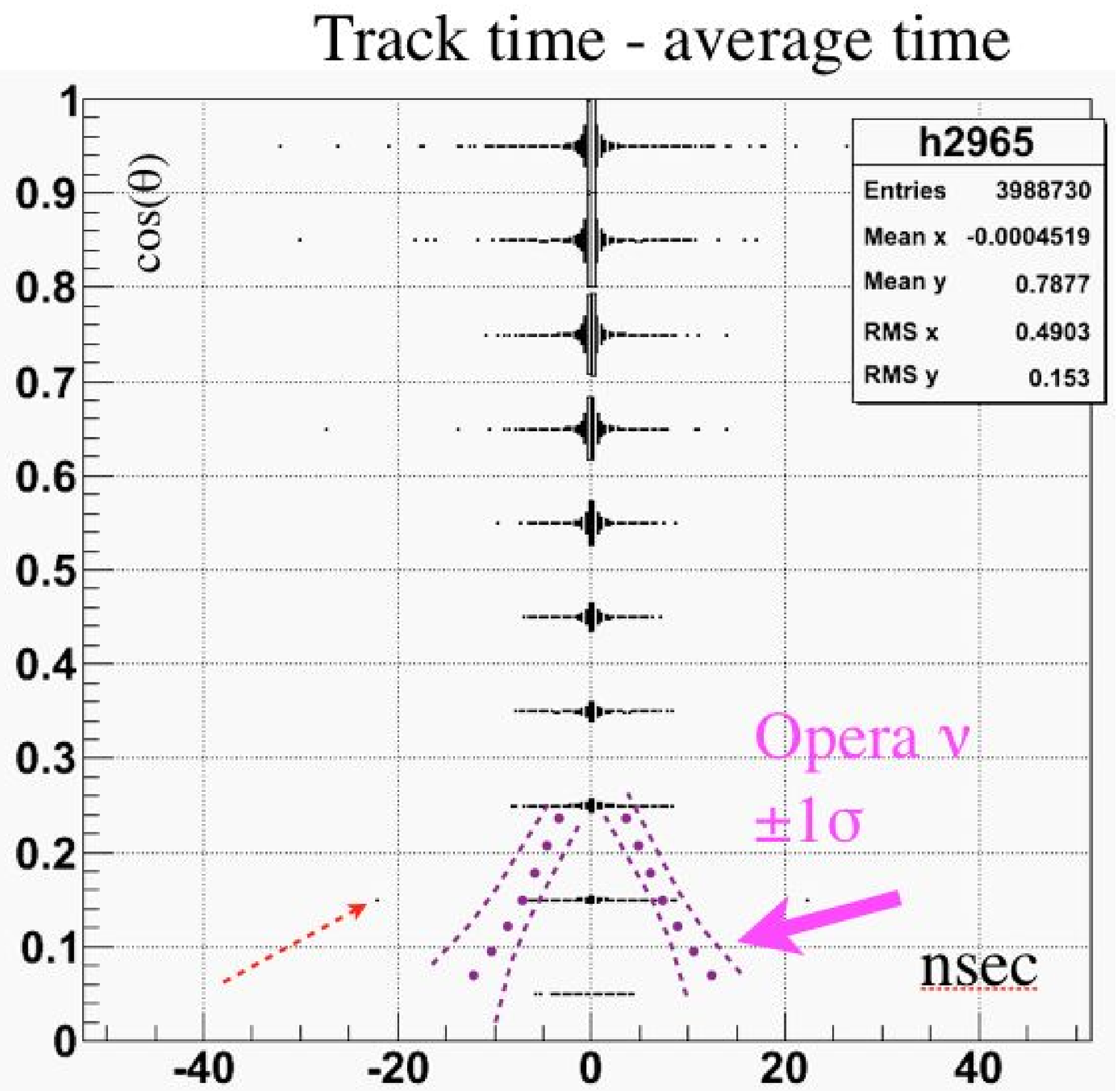}}}
\caption{Difference track time - average time for multiple parallel muons with 2 or 3 tracks as function of $cos(\theta)$. The dot size is proportional to the logarithm of  the bin content. If the original OPERA claim would have been correct, a few tracks are expected inside the dashed region (see text).}
\label{author-fig2}
\end{myfigure}

\section{Time differences in the MACRO muon bundles}

Thanks to its large area and fine tracking granularity  the MACRO detector
was a proper tool for the study of multiple  parallel muons. 
Many papers were published by MACRO on this topic to study the muon multiplicity, the distance between muons and 
the impact on cosmic ray composition from the multiple muon measurement. The last MACRO paper on this argument is in
(Ambrosio, 1999), one important number to consider is that the average distance between muon pairs is $<r>\sim9.4$ m. for vertical tracks and average depth 3800 $gr /cm^2$. The value of $<r>$ changes slowly with depth and zenith angle.
Fig. 1 shows a typical multiple muon event. From this pictures is easy to understand one of the problems of this analysis: the large dimension of the scintillator boxes compared to the average value of $<r>$. The probability to have more than one track intercepting the same counters is high. In this case the time could be wrong. This is because the analysis software could fail
to compute the light propagation time from the intercept of the track to the photomultipliers.
This is the second source of not gaussian tails in the time distributions (the first is the timing between different supermodules).

For each track the analysis program computes the $\beta=v/c$ and the "track time" (average between the scintillator times along the track). To remove noise the analysis program uses only the scintillators in which the position along the scintillator computed by the time differences between the PM at the ends is in agreement with the position given from  the streamer chambers.  Therefore the analysis program computes the differences between the "track times" and the average time of all the tracks in the bundle. This is done including a correction due to  the incidence angle. A $5^{\circ}$ angular cut is applied  to require parallel tracks. To have a valid track time a single scintillator is sufficient but in case of events between different supermodules there is the requirement that at least one track between two different supermodules has two scintillators and that
the beta value is consistent with one. This is to reduce the noise due to the inter ERP TDCs.

\begin{myfigure}
\centerline{\resizebox{90mm}{!}{\includegraphics{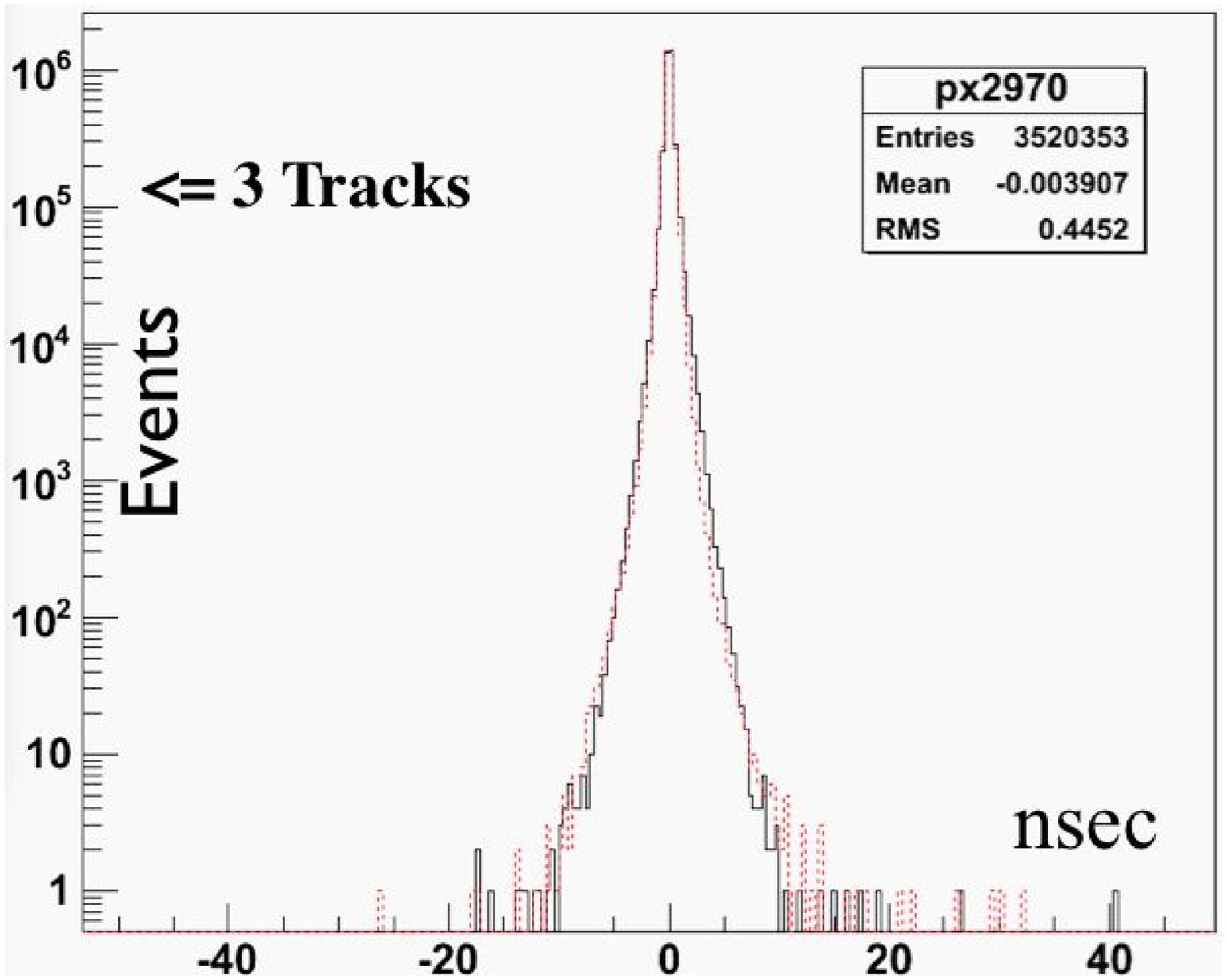}}}
\caption{Difference track time - average time for multiple  muons with 2 or 3 tracks (continuos line) compared with a "simulation" using the data (dashed line)}
\label{author-fig2}
\end{myfigure}

The calculation of the expected number of events, if the original OPERA claim would have been correct,  is done considering the probability to have a neutrino and a muon from the same decay, computed in (Montaruli and Ronga, 2011) and the probability to have a neutrino and a muon from different decays, computed using the approximated Elbert formulas (Gaisser, 1990). The detector and analysis efficiencies have been evaluated using the standard multiple muon MACRO simulation software, with a modification to allow a delay in one of the muons. This calculation gives 2 delayed tracks with time delay $ |\delta t |\ge 10$ nsec expected in MACRO the data set.

In case of events with a muon from a neutrino interaction  is unlikely to have more than one muon directly from the hadronic cascade, so the analysis is limited to events with less than 3 tracks  (one track could be  a spurious track). The results are in Fig 2. Fig 2. also shows the times expected if the original Opera result would have been correct. Considering the region with $cos\theta)\le0.2$ and $ |\delta t |\ge 10$ nsec there is one event with two tracks with a time track - average time ~ 22 nsec  (the dot of Fig 2 near the dashed arrow). However this time is outside the Opera region.  In the Opera region there are no tracks. This result should be compared with the 2 tracks expected.

To understand if the distribution tails in the full angular region  are real or due to detector effects I have done a comparison, computing for each track with two scintillator counters the time difference between times (instead of the average). This is shown in Fig 3 as dashed line. This plot shows that there is agreement between the two distributions and therefore we can conclude that the most of the tail are indeed due to detectors effect.


Finally Fig. 4 shows the time difference including all the multiple muon multiplicity. Since a possible signal due to massive particles or exotic relics is expected at high  path length, I have divided the angular region in two parts: $cos(\theta)\ge.5$, and $cos(\theta)\le.5$. The two distributions are comparable. But  the $cos(\theta)\le.5$ distribution has two tracks with time differences $ \ge15$ nsec, compared with 0.4 tracks  expected from the $cos(\theta)\ge.5$  distribution (Poisson probability 0.06).


\begin{myfigure}
\centerline{\resizebox{90mm}{!}{\includegraphics{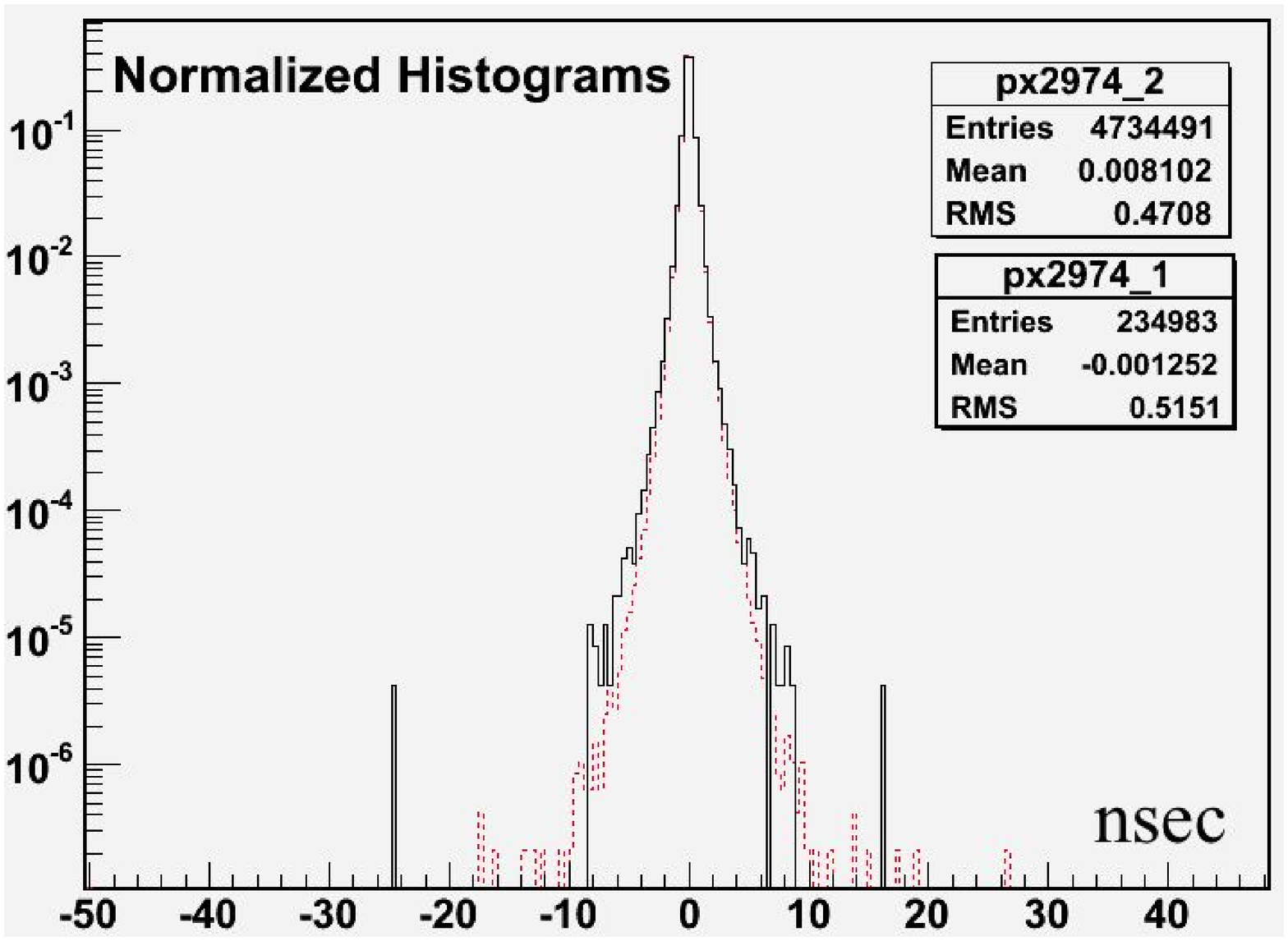}}}
\caption{Difference track time - average time for multiple  muons with all multiplicities: continuous line $cos(\theta)\ge.5$, dashed line $cos(\theta)\le.5$ (histograms normalized to 1).}
\label{author-fig2}
\end{myfigure}

\section{Conclusions}
This work ended some time after the solution of the superluminal neutrino puzzle, however I think that has been very useful to remember that cosmic rays are still important tools in particle physics. For the superluminal neutrino 2 tracks were expected 0 were found.  Considering the different mean-lifes of the pion and the kaon, an "exotic" limit can be derived from  the horizontal tracks of Fig 2 on the equality of the pion and kaon speed in a cascade produced by a primary with $E\ge 3TeV$: 
 $|\beta_{\pi}-\beta_{k}| \ll 1.5 \times 10^{-4}$.
This result is at the moment of very low interest but the superluminal neutrino saga has shown that nothing could be given as guaranteed.
More investigations are necessary on the  delayed tracks in events with multiplicity bigger than 3  and on  massive particles in cosmic rays.

This work has shown once again the importance to save past experiment data for further analysis.  I must thank the  MACRO collaboration that built and run the detector and many MACRO peoples that helped me to recover data and programs and particularly Nazareno Taborgna of the Gran Sasso laboratory, that has been able to save a working alphaVAX with several MACRO original disks. A particular thank is to Teresa Montaruli for useful and deep discussions.

\end{document}